\documentclass[conference]{IEEEtran}
\IEEEoverridecommandlockouts

\usepackage{cite}
\usepackage{amsmath,amssymb,amsfonts}
\usepackage{algorithmic}
\usepackage{graphicx}
\usepackage{textcomp}
\usepackage{xcolor}

\def\BibTeX{{\rm B\kern-.05em{\sc i\kern-.025em b}\kern-.08em
    T\kern-.1667em\lower.7ex\hbox{E}\kern-.125emX}}
\begin{document}

\title{
    CMANet: Channel-Masked Attention Network for Cooperative Multi-Base-Station 3D Positioning\\
    \thanks{
        This work was supported in part by
        National Natural Science Foundation of China under Grant 62071423,
        the Top-Notch Young Talent of China,  Natural Science Foundation of Zhejiang Province under Grant LR23F010004,
        the Top-Notch Young Talent of Zhejiang Province, Fundamental Research Funds for the Central Universities under Grant 226-2024-00125,
        the China Postdoctoral Science Foundation under Grant 2024M762827,
        the Postdoctoral Fellowship Program of CPSF under Grant GZB20230654 and GZC20241521,
        the Postdoctoral Science Preferential Funding of Zhejiang Province ZJ2024054 and ZJ2024055,
        and the Key Research and Development Program of Zhejiang Province under Grant 2024C01160.
    }
}

\author{
    \IEEEauthorblockN{
        Tong An\textsuperscript{$*$},
        Huan Lu\textsuperscript{$*$},
        Jiayang Shi\textsuperscript{$*$},
        Kai Yu\textsuperscript{$\dagger$},
        Rongrong Zhu\textsuperscript{$\ddagger$},
        Bin Zheng\textsuperscript{$*$},
        Jiwei Zhao\textsuperscript{$*$},
        Haibo Zhou\textsuperscript{$\dotplus$},\\
        \textsuperscript{$*$}College of Information Science \& Electronic Engineering, Zhejiang University, Hangzhou, China\\
        \{antoon, luhuan123, jiayangshi, zhengbin, jackokie\}@zju.edu.cn\\
        \textsuperscript{$\dagger$}KTH Royal Institute of Technology, Stockholm, Sweden\\
        kayu@kth.se\\
        \textsuperscript{$\ddagger$}School of Information \& Electrical Engineering, Hangzhou City University, Hangzhou, China\\
        rorozhu@zju.edu.cn\\
        \textsuperscript{$\dotplus$}School of Electronic Science \& Engineering, Nanjing University, Nanjing, China\\
        haibozhou@nju.edu.cn\\
    }
}

\maketitle

\begin{abstract}
    Achieving ubiquitous high-accuracy localization is crucial for next-generation wireless systems, yet remains challenging in multipath-rich urban environments.
    By exploiting the fine-grained multipath characteristics embedded in channel state information (CSI), more reliable and precise localization can be achieved.
    To address this, we present CMANet, a multi-BS cooperative positioning architecture that performs feature-level fusion of raw CSI using the proposed Channel Masked Attention (CMA) mechanism.
    The CMA encoder injects a physically grounded prior—per-BS channel gain—into the attention weights, thus emphasizing reliable links and suppressing spurious multipath.
    A lightweight LSTM decoder then treats \emph{subcarriers as a sequence} to accumulate frequency-domain evidence into a final 3D position estimate.
    In a typical 5G NR-compliant urban simulation, CMANet achieves \textbf{$<$ 0.5\,m} median error and \textbf{$<$ 1.0\,m} 90th-percentile error, outperforming state-of-the-art benchmarks.
    Ablations verify the necessity of CMA and frequency accumulation.
    CMANet is edge-deployable and exemplifies an Integrated Sensing and Communication (ISAC)-aligned, cooperative paradigm for multi-BS CSI positioning.
\end{abstract}


\section{Introduction}
Accurate positioning is a cornerstone for enabling transformative applications in next-generation wireless networks, such as autonomous driving and smart cities.
Not only does it enhance user experience but also plays a critical role in improving network efficiency and resource management\cite{b1, b7, b17, b18}.
However, achieving such high-fidelity positioning, particularly within the complex propagation environments of dense urban settings, remains a significant scientific hurdle.
Conventional satellite-based systems like the Global Positioning System exhibit inherent vulnerabilities; signal obstruction and severe multipath propagation frequently compromise accuracy and reliability, mandating the development of robust terrestrial alternatives\cite{b2}.

Among these alternatives, wireless Channel State Information (CSI), intrinsically available in modern Orthogonal Frequency Division Multiplexing (OFDM) systems, has emerged as an exceptionally promising modality.
CSI provides a fine-grained characterization of the radio channel, capturing frequency-dependent amplitude and phase variations across subcarriers.
This informational richness allows CSI to function as a distinctive spatial signature, offering substantially greater detail than traditional metrics like Time Difference of Arrival, Angle of Arrival, or Received Signal Strength\cite{b21, b22, b23, b24}.
Furthermore, extending this paradigm to incorporate CSI from multiple, spatially distributed base stations (BSs) introduces critical spatial diversity.

\begin{figure*}
    \centerline{\includegraphics[trim=0 0 30 30, scale=0.45]{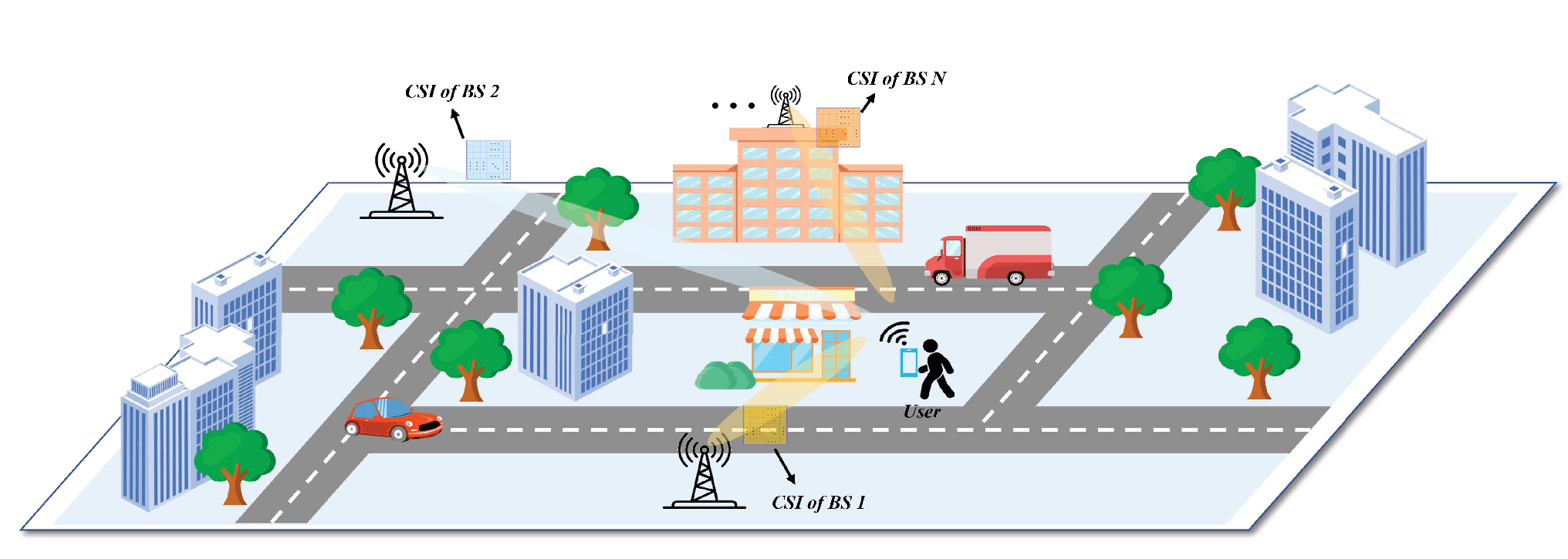}}
    \caption{Multi-base stations joint positioning scenario}
\end{figure*}

Acknowledging CSI's potential, research has explored various positioning algorithms.
Many research efforts focus on transforming CSI into sparse fingerprint features, which are then used as inputs to deep neural networks for localization\cite{b4,b9,b5,b6}.
For instance, reference \cite{b9} proposes converting CSI into an angle-delay channel power matrix (ADCPM) and employs a SegNet combined with MLP for user localization.
References \cite{b5} and \cite{b6} utilize discrete Fourier transforms to convert CSI into usable angle-domain information, as inputs to designed neural networks for localization.
The above fingerprint-based localization schemes enhance CSI's spatial features by constructing sparse matrices; however, much potentially useful information remains unexploited, which could lead to suboptimal localization performance.
So there is an urgent need to develop more effective CSI feature extraction and utilization methods.

For multi-BSs cooperation, existing strategies \cite{b11,b12,b13,b14} frequently prioritize known BS geometric configurations or employ late-stage fusion, aggregating position estimates derived independently from each BS.
A critical limitation is their failure to fully harness the rich, synergistic information embedded within the raw CSI measurements obtained concurrently from multiple network vantage points.
Specifically, the intricate inter-channel dependencies and fine-grained correlations—how the complex multipath signature observed at one BS relates to and complements signatures at others—remain relatively under-explored and suboptimally utilized.
This oversight constitutes significant untapped potential, likely constraining achievable localization accuracy and restricting robust performance primarily to less complex or geographically limited contexts.

To decisively address this identified gap and unlock the latent capabilities of distributed CSI sensing, this paper introduces a novel multi-base station joint positioning network called CMANet.
It operationalizes this philosophy via a meticulously architected end-to-end framework comprising two integral, synergistically interacting modules.
The encoder centers around a well-designed amplitude masked attention block, which adaptively modulates attention weights by integrating CSI amplitude features, thereby extracting globally relevant features from the CSI data.
The decoder is designed to cumulatively process frequency-domain features across subcarriers, effectively capturing temporal dependencies and refining position estimates.
This bespoke hybrid architecture enables sophisticated hierarchical spatio-temporal dependency modeling across the distributed multi-BS infrastructure.
The principal contributions of this research are as follows:
\begin{itemize}
    \item \textbf{Native CSI fusion paradigm:} We propose the first feature-level early fusion of \emph{raw} multi-BS CSI, explicitly modeling cross-BS multipath complementarity, in contrast to late fusion of per-BS estimates or heavy hand-crafted fingerprints.
    \item \textbf{Channel masked attention:} We introduce CMA, which normalizes per-BS channel gains to guide attention weights, focusing learning capacity on reliable links.
    \item \textbf{Frequency-sequence accumulation:} We treat subcarriers as a sequence and use a well-designed recurrent neural network to aggregate frequency-domain evidence, explaining the observed error reduction with more subcarriers.
    \item \textbf{Performance improvement and deployability:} In a 5G NR urban scenario, CMANet attains sub-meter accuracy at the 90th percentile while maintaining good scalability that is suitable for varying numbers of base stations and subcarriers.
\end{itemize}

\section{System Model}
In this section, we first present the signal model of the OFDM system. Then, we introduce the learning-based CSI positioning framework.
\begin{figure*}[t]
    \centerline{\includegraphics[width=\textwidth]{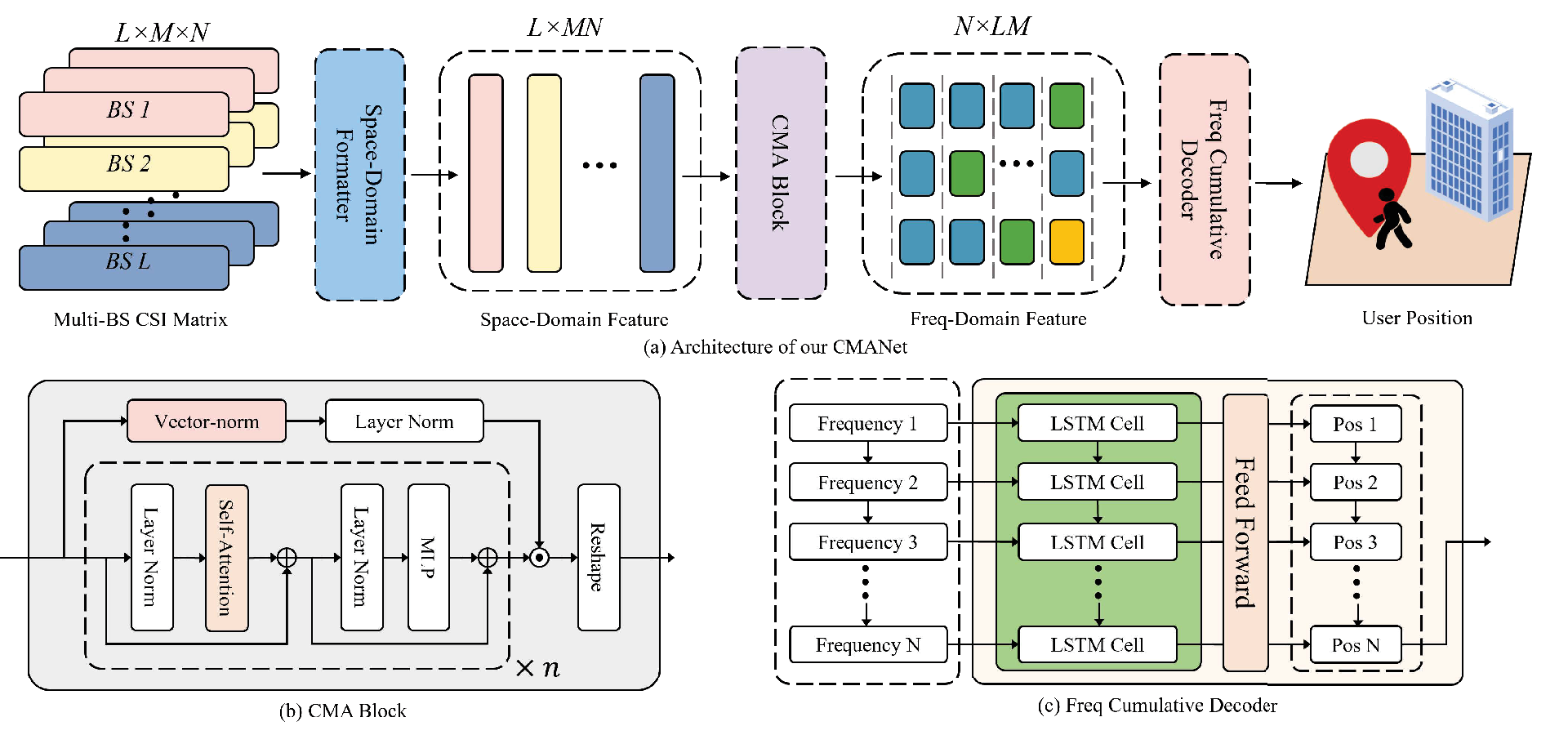}}
    \caption{The architecture of CMANet based on the channel masked attention}
    \label{CMANet Architecture}
\end{figure*}
\subsection{OFDM Channel Model}
We consider an OFDM communication system with $M$ antennas. The wireless channel between a user and a base station is modeled as having $P$ propagation paths, and the frequency-domain channel response at frequency $f$ is expressed as
\begin{equation}
    R\left(f\right)=\sum_{p=1}^{P}{a_pe^{j\varphi_p}\bf{a}\left(\theta_p\right)}
\end{equation}
where $a_p$ and $\varphi_p$ represent the amplitude gain and phase offset of the $p$-th propagation path, respectively.
$\bf{a}\left(\theta_p\right)$ denotes the array steering vector corresponding to the angle of arrival $\theta_p$, defined as
\begin{equation}
    \bf{a}\left(\theta_\textit{p}\right)=\left[\textit{e}^{-\textit{j}\frac{2\pi \textit{d}_\textit{1}}{\lambda}cos{\theta_\textit{p}}},\ \cdots,\textit{e}^{-\textit{j}\frac{2\pi \textit{d}_\textit{M}}{\lambda}cos{\theta_\textit{p}}}\right]^T
\end{equation}
where $d_m$ is the propagation distance between user and the $m$-th antenna, $\lambda=c/f$ is the wavelength, $c$ is the speed of light and $f$ is the frequency of subcarrier.
For an OFDM system with $N$ subcarriers, the channel state matrix between the user and a base station is given by
\begin{equation}
    I=\left[R\left(f_1\right),\ R\left(f_2\right),\ \cdots,\ R\left(f_N\right)\right]
\end{equation}
where each subcarrier frequency is defined as
\begin{equation}
    f_k=f_c+\left(k-N/2\right)\cdot \Delta f
\end{equation}
Here, $f_c$ is the carrier frequency and $\Delta f$ is the subcarrier spacing. When the user is served by $L$ base stations, the overall channel state information can be stacked to form a tensor
\begin{equation}
    H=\left[I_1,\ I_2,\ \cdots,\ I_L\right]^T
\end{equation}
This composite matrix $H$ forms the foundation of our input data for neural network-based localization.

\subsection{Neural Network-Based Position Estimation}
The joint CSI matrix ($H$) which captures amplitude and phase information across all subcarriers from multiple base stations, is used as the input to our deep learning model for position estimation.
We denote the output position estimate as
\begin{equation}
    \hat{x}=F\left(H\right)
\end{equation}
where $\hat{x}\in\mathbb{C}^{3}$ is the predicted 3D spatial coordinate of the user, and $F\left(\cdot\right)$ s a function representing the trained neural network.

The learning-based localization framework operates in two main phases: training phase and deployment phase.
In training phase, historical CSI data and corresponding user position labels are collected from multiple base stations.
These paired datasets are used to train the network parameters via supervised learning.
Once trained, the model can infer the user's position in real-time using newly acquired CSI data without requiring additional geometric information or manual feature engineering.

\section{Multi-Base Stations Joint Positioning Network}
To enable high-precision localization in complex outdoor environments, we propose a cooperative positioning framework that jointly leverages CSI from multiple base stations.
This section presents the architecture of the proposed CMANet, which utilizes a hybrid deep learning approach to model spatial and frequency dependencies across multi-BS CSI data.

As illustrated in Fig. \ref{CMANet Architecture}(a), the CMANet framework consists of three core modules: Space-Domain Format Module, Channel Masked Attention Block, and Frequency Cumulative Decoder.

\subsection{Space-Domain Format Module}
The purpose of this module is to transform the input complex CSI tensor into a format suitable for the attention module.
Considering that the original CSI matrix is in complex format, we must decompose it into a real-valued format that is suitable for input into a neural network.
This is accomplished by separating the real and imaginary parts to construct an equivalent tensor $H_1\in\mathbb{C}^{L\times 2M\times N}$ formulated as
\begin{subequations}
    \begin{equation}
        H_1\left[L,2m,N\right]=Re\left(H\left[L,m,N\right]\right)
    \end{equation}
    \begin{equation}
        H_1\left[L,2m+1,N\right]=Im\left(H\left[L,m,N\right]\right)
    \end{equation}
\end{subequations}

Then, we flatten it along the antenna and subcarrier dimensions to form an input tensor with a sequence length equal to the number of base stations and an embedding dimension equal to the number of antennas times subcarriers, expressed as
\begin{equation}
    H_1\in\mathbb{R}^{L\times 2M\times N}\xrightarrow{\textit{flatten}} H_2\in\mathbb{R}^{L\times2MN}
\end{equation}

The rationale behind this design is that CSI data from different base stations are spatially distributed, and each base station's CSI data contains rich spatial information.
Therefore, by treating the base station dimension as the sequence length and the product of antenna and subcarrier dimensions as the embedding dimension, we create a serialized input tensor.
This approach enables the attention module to effectively capture spatial correlations and temporal features across base stations.
Additionally, this design offers scalability and adaptability; increasing the number of base stations simply involves extending the sequence length dimension of the input tensor without altering the embedding dimension.

\subsection{Channel Masked Attention Block}
The channel masked attention is central to capturing interdependencies within the formatted CSI input.
Considering the varying contributions of different base stations to the positioning results, we designed the masked attention with channel gain as the weighting factor to achieve more reasonable perception results.
Present in Fig. \ref{CMANet Architecture}(b), the process is expressed as follows:
\begin{subequations}
    \begin{equation}
        C_{gain} = VectorNorm\left(H_2\right)
    \end{equation}
    \begin{equation}
        W = LayerNorm\left(C_{gain}\right)
    \end{equation}
    \begin{equation}
        H_3 = W \odot SelfAttention(H_2)
    \end{equation}
\end{subequations}
where $C_{gain}$ presents the channel gain vector obtained by calculating the vector norm of each base station's CSI data,
$W$ is the attention weight vector derived from normalizing the channel gain vector using layer normalization.
To be note that $W\in\mathbb{R}^{L\times 1}$ and $H_2\in\mathbb{R}^{L\times 2MN}$, so the element-wise multiplication $\odot$ is performed by broadcasting $W$ along the embedding dimension.
$H_3\in\mathbb{R}^{L\times 2MN}$ is the output of the self-attention mechanism weighted by the channel gain.

This module is designed to account for the varying influence of different base stations' channel gains on the positioning results.
By doing so, the attention can be directed towards base stations with higher channel gains, thereby enhancing the accuracy and robustness of the positioning.

At the output layer of the module, we include a reshape operation, which involves unflattening, permuting and flatten, to adjust the shape of the output tensor to be suitable for input into the LSTM-Based Decoder module, expressed as
\begin{subequations}
    \begin{equation}
        H_3\in\mathbb{R}^{L\times 2MN}\xrightarrow{\textit{unflatten}} H_4\in\mathbb{R}^{L\times 2M\times N}
    \end{equation}
    \begin{equation}
        H_4\in\mathbb{R}^{L\times 2M\times N}\xrightarrow{\textit{permute}} H_5\in\mathbb{R}^{N\times 2M\times L}
    \end{equation}
    \begin{equation}
        H_5\in\mathbb{R}^{N\times 2M\times L}\xrightarrow{\textit{flatten}} H_6\in\mathbb{R}^{N\times 2ML}
    \end{equation}
\end{subequations}

\subsection{Frequency Cumulative Decoder}
From the output tensor of the encoding module, we observe that its shape is (subcarriers, antennas $\times$ BSs), where subcarriers presents frequency-domain information, and the product of antennas and BSs presents spatial information.
To leverage this rich spatial information, we treat the subcarrier dimension as the temporal dimension and employ an LSTM network to progressively decode the user's spatial location, shown in Fig. \ref{CMANet Architecture}(c).
Along the subcarrier dimension, each set of spatial features is decoded by the LSTM cells described in \cite{b20}. Forwarding a parameters-sharing multi-layer perceptron (MLP), the spatial features are mapped to user's spatial coordinates.
The positioning result, which integrates information from all subcarriers, is expressed as:
\begin{equation}
    \hat{x} = MLP\left(LSTM\left(H_6\right)\right)
\end{equation}
where $\hat{x}\in \mathbb{R}^{N\times3}$ presents each subcarrier's position estimate, and the final output is the last subcarrier's position estimate, expressed as:
\begin{equation}
    \hat{x}_{final} = \hat{x}\left[N\right]
\end{equation}

\subsection{Loss Function}
To train the network, we minimize the Weighted Mean Squared Error (W-MSE) between the predicted and ground truth positions
\begin{equation}
    \mathcal{L} = \sum_{i=1}^{N}\frac{i}{N}\left\|x_i-\hat{x}\left[N\right]\right\|_2
\end{equation}
where $N$ is the number of training samples, $x_i$ is the ground truth position of the $i$-th sample.

The rationale behind this loss function design is to guide the model to focus more on the user's final position by increasing the loss weight for later subcarriers, thereby improving positioning accuracy.
Additionally, for optimizing the model parameters, mainstream optimization algorithms such as Adam or SGD can be employed.

\section{Simulation Results}
This section presents the simulation settings and performance evaluations of the proposed CMANet framework.
We conduct extensive experiments in realistic outdoor environments to validate the effectiveness of our approach and benchmark it against state-of-the-art algorithms.

\subsection{Parameters Settings}
We simulate a representative urban environment near the Arc de Triomphe in Paris, using open-source geographic data from OpenStreetMap.
The 3D urban scene is modeled and rendered using Blender in combination with the Mitsuba-Blender add-on to generate geometry files compatible with the Mitsuba 3 rendering engine.
The ray tracing is conducted using Sionna, a wireless simulation library developed by NVIDIA\cite{b16}.
In this simulated scene, six base stations (BSs) with identical configurations are deployed to provide wireless coverage and CSI acquisition.
The placement of BSs reflects real-world base station locations obtained from OpenCelliD data\cite{b19}.
User equipment (UE) is assumed to be randomly distributed within a rectangular area of 220 m $\times$ 300 m.
The UE device height is randomly distributed between 0-30m, simulating ground user or aerial drone positioning in urban scenarios.
The simulation parameters—compliant with 3GPP specifications—are listed in Table \ref{SIMULATION PARAMETERS}.
\begin{table}[h]
    \caption{SIMULATION PARAMETERS}
    \begin{center}
        \begin{tabular}{p{4.5cm}|p{2.5cm}}
            \hline
            \textbf{Parameters}        & \textbf{Values} \\
            \hline
            Number of BS$(L)$          & 6               \\
            \hline
            Transmission Direction     & Uplink          \\
            \hline
            Center frequency$(f_c)$    & 3.5GHz          \\
            \hline
            Bandwidth                  & 20MHz           \\
            \hline
            Number of Subcarriers$(N)$ & 288             \\
            \hline
            Antenna Array Form         & Rectangular UPA \\
            \hline
            Rows of Antennas           & 2               \\
            \hline
            Columns of Antennas        & 4               \\
            \hline
            Spacing of Antennas$(d)$   & 0.5$\lambda_c$  \\
            \hline
        \end{tabular}
    \end{center}
    \label{SIMULATION PARAMETERS}
\end{table}

\subsection{Performance Assessment}
For performance assessment, we consider the following compared algorithms:
\begin{itemize}
    \item \textbf{Channel Masked Attention Model}: The proposed CMANet cooperative architecture described in Fig. \ref{CMANet Architecture}.
    \item \textbf{Self-Attention Model}: The proposed CMANet architecture with no channel masking but fully self-attention described in \cite{b15}.
    \item \textbf{ADCPM-Based Cooperative DL model}: A cooperative architecture taking in ADCPM fingerprints as input and using SegNet-MLP for localization, referenced in \cite{b9}.
    \item \textbf{MFCNet}: A deep learning-based positioning algorithm designed for OFDM networks, referenced in \cite{b3}.
\end{itemize}

\begin{figure}[h]
    \centerline{\includegraphics[trim=0 0 0 10, scale = 0.52]{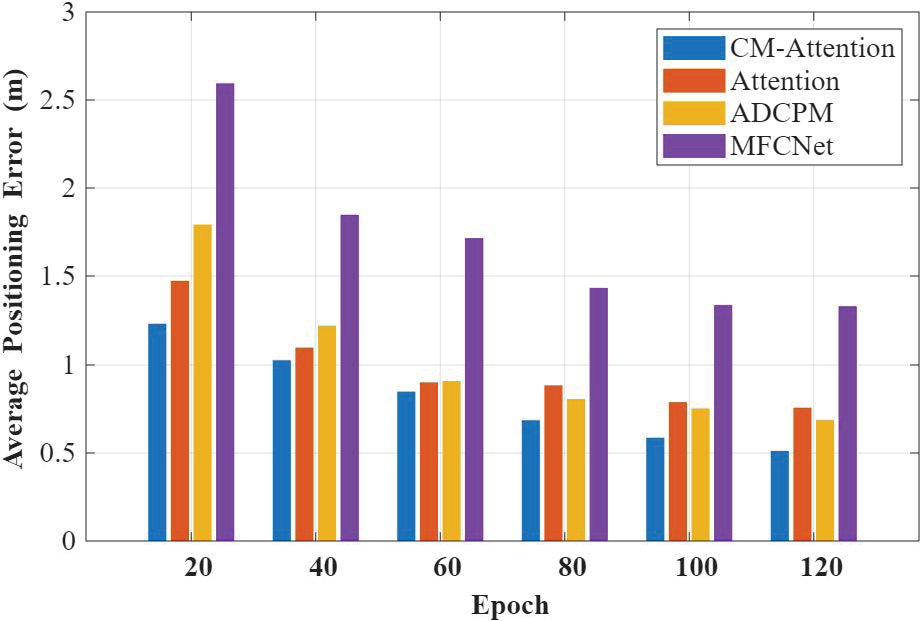}}
    \caption{Positioning Accuracy During Training}
    \label{Positioning Accuracy During Training}
\end{figure}

\textit{a) Positioning Accuracy:}
Positioning accuracy is a critical metric for evaluating the performance of localization algorithms.
We first compare the performance of the proposed algorithm with the benchmarks.
During the training phase, in each epoch, we randomly generate 10,000 different topologies within a rectangular area of 220 m $\times$ 300 m, with heights randomly distributed between 0-30m.
Ray tracing is performed using Sionna to generate CSI data from six base stations for each topology.
Every 20 epochs, we randomly generate 1,000 different topologies within the same area as the test set to evaluate the model's positioning accuracy.
The model's positioning performance is measured by calculating the positioning error, defined as the Euclidean distance between the predicted position and the true position, shown in Fig. \ref{Positioning Accuracy During Training}.
The figure shows that the positioning error of all algorithms decreases with the increase of training epochs, indicating that the models are continuously learning and optimizing.
Among them, the proposed algorithm (blue bar) exhibits a faster convergence rate and lower final positioning error during training, demonstrating its advantages in multi-base station cooperative positioning.

\begin{figure}[h]
    \centerline{\includegraphics[trim=0 0 0 10 ,scale = 0.52]{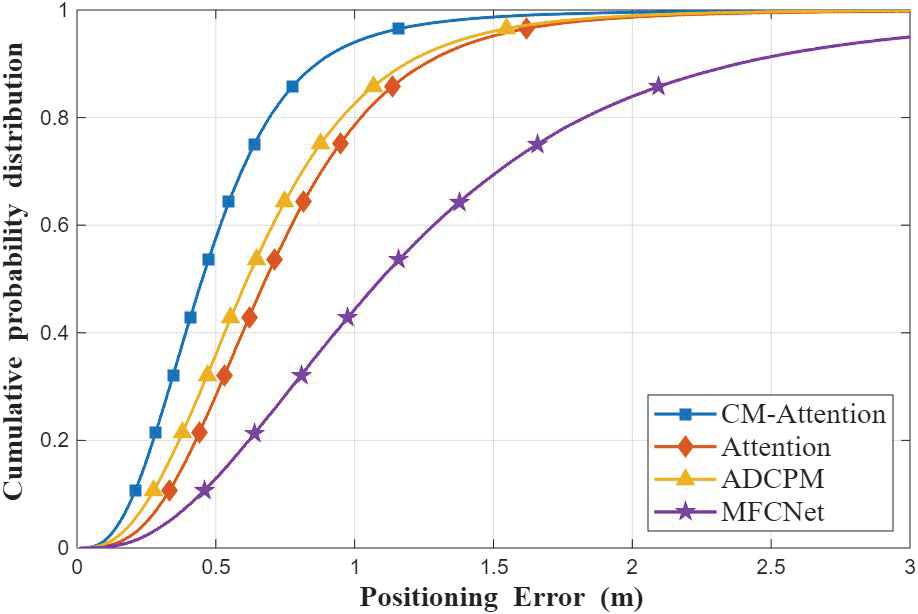}}
    \caption{CDF of Different Algorithms}
    \label{CDF}
\end{figure}

After 140 training epochs, we deploy the four trained models in the corresponding environment and generate an additional 20,000 different topologies for testing.
Fig. \ref{CDF} presents the cumulative distribution function (CDF) of positioning errors for each algorithm on the test set.
Notably, the pure self-attention model without the masking module (orange curve) underperforms the ADCPM-based cooperative DL model (yellow curve) in terms of positioning error.
The introduction of the masking module (blue curve) significantly enhances positioning performance, indicating that channel gain information plays a crucial role in multi-base station cooperative positioning.
Additionally, the proposed algorithm achieves positioning errors of less than 0.5m for 50\% of the cases and less than 1.0m for 90\% of the cases, outperforming all other benchmark algorithms.

\begin{figure}[b]
    \centerline{\includegraphics[trim=0 0 0 10, scale = 0.52]{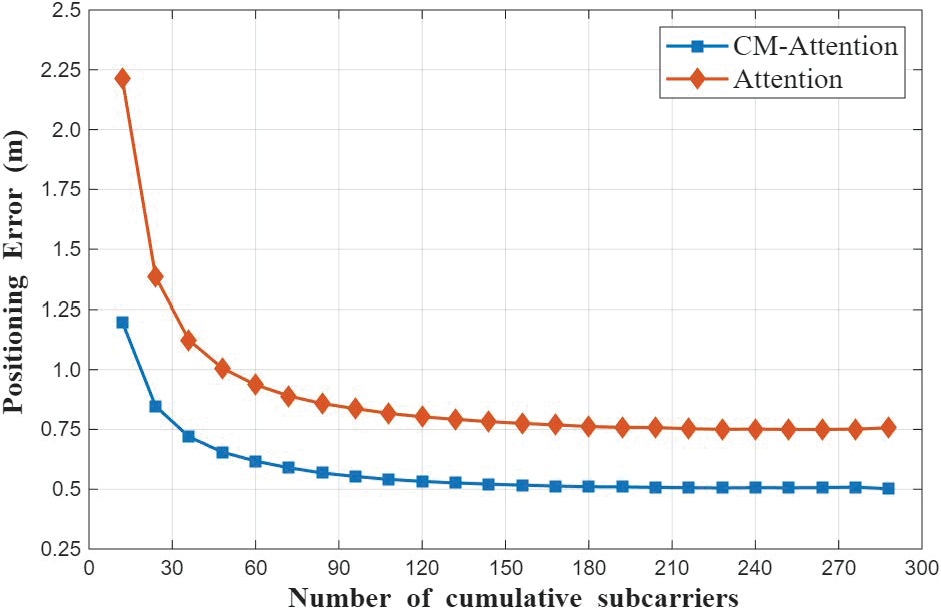}}
    \caption{Cumulative Effect of Frequency Domain Features}
    \label{Cumulative Effect of Frequency Domain Features}
\end{figure}

\begin{figure*}
    \centerline{\includegraphics[trim=0 0 0 0, width=\textwidth]{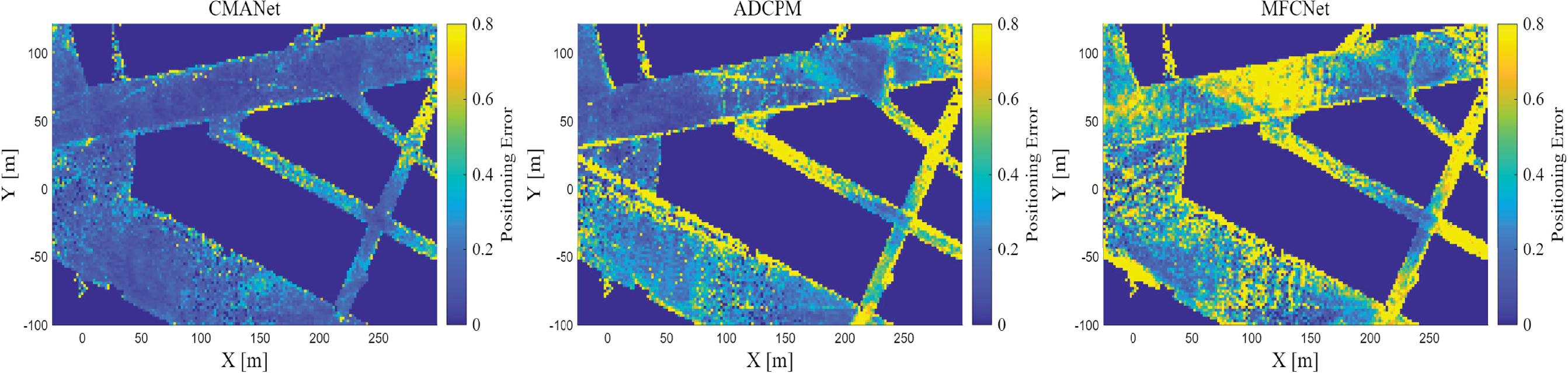}}
    \caption{Error Hotspot Map of CMANet and ADCPM-Based Model}
    \label{Error Hotspot Map}
\end{figure*}

\textit{b) Cumulative Effect of Frequency Domain Features:}
To further illustrate the cumulative effect of frequency domain features on positioning accuracy, we present the positioning error variation of CMANet by cumulatively increasing the number of subcarriers.
Similarly, we generate an additional 20,000 different topologies for testing, recording the positioning error every 12 subcarriers (1 resource block), as shown in Fig. \ref{Cumulative Effect of Frequency Domain Features}.
The figure shows that as the number of cumulative subcarriers increases, the positioning error gradually decreases.
This phenomenon shows that by ensembling more subcarrier information, CMANet obtains a more accurate position inference result, which intuitively shows the multi-frequency ensemble effect, and also validates the rationality of the designed network structure.

\textit{c) Error Hotspot Map:}
We plot the positioning error heatmaps, as shown in Fig. \ref{Error Hotspot Map}, to visually demonstrate the positioning performance of the proposed algorithm at different geographic locations.
In these heatmaps, areas closer to yellow indicate larger positioning errors, while areas closer to blue indicate smaller positioning errors.
We observe that the proposed algorithm achieves low positioning errors in most areas, even in some complex urban environments, the positioning errors remain within a small range.
In contrast, the ADCPM-based cooperative DL model exhibits larger positioning errors in certain areas, especially in regions with weaker base station coverage or significant multipath effects.

\section{CONCLUSION}
In this work, we have introduced CMANet, a novel deep learning framework for cooperative positioning using CSI from multiple OFDM base stations, showcasing its potential as a key enabler for 6G ISAC systems.
The proposed architecture pioneers a new paradigm for cellular localization by combining native feature cooperation, space-aware attention mechanisms, and frequency-domain sequence accumulation.
These innovations address the critical limitations of single-base station systems, making CMANet particularly suitable for high-precision positioning in vehicular networks, drone navigation, and outdoor robotics.
Future work will focus on enhancing CMANet with online adaptation capabilities, cross-city generalization to improve its applicability and reliability.

\end{document}